**Testing for the Asymmetric Optimal Hedge Ratios: With an Application to Bitcoin**

Abdulnasser Hatemi-J

Department of Economics and Finance, College of Business and Economics, UAE University

Email: AHatemi@uaue.ac.ae

Abstract

Reducing financial risk is of paramount importance to investors, financial institutions, and corporations. Since the pioneering contribution of Johnson (1960), the optimal hedge ratio based on futures is regularly utilized. The current paper suggests an explicit and efficient method for testing the null hypothesis of a symmetric optimal hedge ratio against an asymmetric alternative one within a multivariate setting. If the null is rejected, the position dependent optimal hedge ratios can be estimated via the suggested model. This approach is expected to enhance the accuracy of the implemented hedging strategies compared to the standard methods since it accounts for the fact that the source of risk depends on whether the investor is a buyer or a seller of the risky asset. An application is provided using the spot and the futures prices of Bitcoin. The results strongly support the view that the optimal hedge ratio for this cryptocurrency is position dependent. The investor that is long in Bitcoin has a much higher conditional optimal hedge ratio compared to the one that is short in the asset. The difference between the two conditional optimal hedge ratios is statistically significant, which has important repercussions for implementing risk management strategies.

# 1. Introduction

One of the main functions of derivative securities is to achieve hedging, which means reducing (or neutralizing if possible) the risk of the price of an asset via the derivative security. Futures contracts are regularly used for this purpose. Whether to have a long position or a short position in a futures contract for hedging purposes depends on the need of the hedger. Assume that the relationship between the spot instrument and the futures as the hedging instrument is positive. If the hedger has a long position in the original asset, then the need is to sell in the future and therefore a short position in the futures contract is appropriate in this case. On the other hand, if the hedger needs to buy the original asset in the future, then a long position in the futures is required for dealing with the underlying risk. Another important issue within this context that is usually neglected in the literature is the difference in the source of risk depending on whether the trader is buyer or seller of the original asset. The source of risk for a trader that is short in the risky asset is increasing prices while the source of risk for a trader with a long position is decreasing prices. For example, an oil importing country as a buyer is exposed to increasing oil prices risk whereas an oil exporting country's risk exposure is determined by falling oil prices. What is the good news for one party is the bad news for the other party and *vice versa*. Hence, any price change is not part of the risk for any given trading position and there is a need for a position depend hedging strategy. Since there are potentially two different values for the optimal hedge ratio depending on whether the trader is short or long in the risky asset, it is crucial to assess not only if these estimated values are statistically significant individually, but also if their difference is statistically significant. Thus, this paper aims at introducing an efficient test of difference between the estimated two values that can be implemented by using a system of equations for detecting whether the optimal hedge ratios are asymmetric or not. To the author's best knowledge, this is the first attempt to provide a method for testing the asymmetric optimal hedge ratios, which explicitly accounts for the position the trader is taking in the risky asset. An application to Bitcoin is provided, which is the cryptocurrency that has the largest market capitalization worldwide and it also has a functional futures market.

The rest of this paper is structured as follows. Section 2 provides a brief review of the literature. The method for testing the null hypothesis of symmetric optimal hedge ratio against the asymmetric alternative one is introduced in Section 3. A numerical application on hedging the price risk of Bitcoin is provided in Section 4. The conclusions are offered in the last section.

## 2. A Brief Literature Review

Two influential studies conducted by Dyhrberg (2016 a, b) exhibit empirically that a cryptocurrency can perform as useful hedge for protecting the value of some other investment assets. Będowska-Sójka and Kliber (2022) provide empirical results that show cryptocurrencies can be useful as a hedge again oil price fluctuations to some extent. Shahzad et. al., (2022) investigate the potential hedging performance of Bitcoin for protecting the equity markets of the BRIC countries via dynamic methods. They present empirical results showing that Bitcoin can perform as an efficient hedge for protecting the value of the financial markets in these countries. According to Urquhart and Zhang (2019) Bitcoin can function as a hedge instrument for other currencies based on the empirical findings obtained using intraday data. Chen et. al., (2024) provide a complete literature review of hedge ratios.

The proposed method in this paper is used to test whether or not the futures of Bitcoin can function as a hedge to protect the spot value of Bitcoin from the perspective of both buyer and seller of this very risky asset. Bitcoin was introduced by Nakamoto (2008), which has the highest market value currently compared to many alternative ones.[1] Cryptocurrencies are characterized by illiquidity, structure breaks with alternations and extremely high volatility. Therefore, hedging is of paramount importance when dealing with cryptocurrencies. Currently, the only cryptocurrency that has a futures market is Bitcoin.

## 3. Methodology

In this section symmetric and asymmetric optimal hedge ratios are defined. An efficient test is also introduced to the test the null hypothesis of symmetric optimal hedge ratio against the asymmetric alternative hypothesis. It is shown how the asymmetric optimal hedge ratios can be estimated when needed via a system of equations that is also flexible to have multivariate GARCH effects with a potential multivariate t-distribution.

### 3.1 The Symmetric Optimal Hedge Ratio

---

[1] Bhimani et. al., (2022) provides an interesting study of factors determining the adaptation of cryptocurrencies in 137 countries. Stimulating discussions on advantages and disadvantages of cryptocurrencies are offered by Taleb (2021) and Lipton (2021). For a detailed review of the literature on cryptocurrencies see Bariviera and Merediz-Solà (2021). Erdogan et. al., (2022) investigate the impact of cryptocurrencies on the natural environment. A stochastic difference equation for pricing of cryptocurrencies is introduced in El-Khatib and Hatemi-J (2024). For a recent investigation of the dynamics in the cryptocurrency markets see Yan (2024). Huang (2024) provides new insights on the dynamic interaction between conventional financial markets and the markets for cryptocurrencies. For modelling time-dependent and position dependent measures of Bitcoin volatility see Hatemi-J (2024).

The optimal hedge ratio (OHR), which was introduced by Johnson (1960) originally, aims at making sure that the total value of the hedged portfolio remains unaltered (see Hull, 2012) by having an offsetting position in the futures contract versus the original asset. The main question in this case is how many futures contract is needed for each unit of the original asset that would minimize the risk of the hedged portfolio? The answer to this question can be provided by estimating the OHR. Note that the hedged portfolio includes the quantities of the spot instrument (i.e., the original asset in this case) as well as the hedging instrument (i.e., the futures contract in this case) and it can be expressed mathematically as the following according to Johnson (1960):

$$V_h = Q_s S - Q_f F \tag{1}$$

Where $V_h$ represents the value of the hedged portfolio, $Q_s$ and $Q_f$ are the quantity of spot and futures instrument respectively. $S$ and $F$ signify the prices of the spot and the futures, respectively. Equation (1) can be transformed into changes. Since the sources of uncertainty are the two prices only, equation (1) can be expressed as the following:

$$\Delta V_h = Q_s \Delta S - Q_f \Delta F \tag{2}$$

Where $\Delta$ is the first difference operator. The ultimate goal of the hedging strategy is to achieve $\Delta V_h = 0$, which results in having $\frac{Q_f}{Q_s} = \frac{\Delta S}{\Delta F}$. Now, let us define the following ratio:

$$h = \frac{Q_f}{Q_s} \tag{3}$$

Which implies

$$h = \frac{\Delta S}{\Delta F}. \tag{4}$$

Consequently, $h$ represents the hedge ratio, which can be obtained as the slope parameter in a regression of the price change of the spot instrument ($\Delta S$) on the price change of the futures ($\Delta F$) as the hedging instrument. This point can be demonstrated mathematically. By substituting equation (3) into equation (2) the following expression can be obtained:

$$\Delta V_h = Q_s [\Delta S - h \Delta F] \tag{5}$$

The optimal hedge ratio (OHR) is the one that minimizes the risk of a change in the value of the portfolio that is hedged. This risk is measured by the variance of equation (5), which is given by the following equation:

$$Var[\Delta V_h] = Q_s^2 \left[\sigma_S^2 + h^2 \sigma_F^2 - 2h\rho\sigma_S\sigma_F\right] \qquad (6)$$

Here $\sigma_S^2$ and $\sigma_F^2$ represent the variance of $\Delta S$ and $\Delta F$, respectively. The correlation coefficient between $\Delta S$ and $\Delta F$ is denoted by $\rho$. In order to obtain the OHR, equation (6) needs to be minimized with respect to the parameter $h$. That is,

$$\frac{\partial[Var[\Delta V_h]]}{\partial h} = Q_s^2 \left[2h\sigma_F^2 - 2\rho\sigma_S\sigma_F\right] = 0, \qquad (7)$$

Which gives

$$h^* = \rho \frac{\sigma_S}{\sigma_F}. \qquad (8)$$

This OHR can also be obtained by estimating the following regression model:

$$\Delta S_t = \alpha + h\Delta F_t + u_t, \qquad (9)$$

For $t=1, \ldots, T$. Where $\alpha$ is an intercept and $u_t$ is an error term. The estimated slope in equation (9), i.e. $h$, is the optimal hedge ratio that is exactly equal to the value obtained by equation (8).

It should be mentioned that the sign of the optimal hedge ratio matters. The strategies that are defined in Table 1 describe how the investor should create the hedged portfolio depending on whether the optimal hedge ratio is positive or negative.

Table 1. Strategies for Hedging via Futures.

| When $h > 0$ | When $h < 0$ |
| --- | --- |
| Long original asset + short futures | Long original asset + long futures |
| Short original asset + long futures | Short original asset + short futures |

Note that if the strategies defined in Table 1 are not exactly implemented and an opposite position is undertaken in any case, it is going to lead to increasing risk, which means speculation. Thus, it is very crucial for the investor to make sure that the goal and the strategy are consistent with each other before the implementation of any strategy.

### 3.2 The Asymmetric Optimal Hedge Ratios

The asymmetric optimal hedge ratios have also been introduced in the literature (El-Khatib and Hatemi-J, 2011).[2] Allowing for asymmetry accords well with reality because investors tend to react stronger to negative changes compared to the positive ones. In addition, the asymmetric hedge ratios provide position dependent hedging values.[3] These ratios are denoted by $h^+$ and $h^-$, which can be calculated via the following equations:

$$h^+ = \rho^+ \frac{\sigma_S^+}{\sigma_F^+} \quad \text{or} \quad h^- = \rho^- \frac{\sigma_S^-}{\sigma_F^-}.$$

Here $\sigma_S^+$ signifies the standard deviation of $\Delta S^+$, $\sigma_F^+$ is the standard deviation of $\Delta F^+$ and the correlation coefficient between $\Delta S^+$ and $\Delta F^+$ is denoted by $\rho^+$. Similar values are defined for the negative component. The following equations are used to transform the data into positive and negative components:

$$\Delta S_t^+ = MAX(\Delta S_t, 0) \tag{10}$$
$$\Delta F_t^+ = MAX(\Delta F_t, 0) \tag{11}$$
$$\Delta S_t^- = MIN(\Delta S_t, 0) \tag{12}$$

and

$$\Delta F_t^- = MIN(\Delta F_t, 0) \tag{13}$$

In the next sub-section, an explicit test for the null hypothesis of no asymmetry in the optimal hedge ratio is presented.

---

[2] There have been previous attempts for allowing some form of asymmetry in the volatility for the optimal hedge ratio without explicitly providing position dependent optimal hedge ratios, such as Brooks et. al, (2002), among others.
[3] For stochastic optimal hedge ratios see Hatemi-J and El-Khatib (2012).

## 2.3 Asymmetric Optimal Hedge Ratio Test

In order to detect whether or not the hedge ratio is symmetric, we suggest estimating the following system of regression equations:

$$\Delta S_t^+ = \alpha^+ + h^+ \Delta F_t^+ + u_t^+ \quad (14a)$$

$$\Delta S_t^- = \alpha^- + h^- \Delta F_t^- + u_t^- \quad (14b)$$

$$\sigma_t^{2+} = \gamma^+ + \sum_{i=1}^{K^+} \phi_j^+ u_{t-i}^{2+} + \sum_{j=1}^{Q^+} \lambda_j^+ \sigma_{t-j}^{2+} \quad (14c)$$

$$\sigma_t^{2-} = \gamma^- + \sum_{i=1}^{K^-} \phi_j^- u_{t-i}^{2-} + \sum_{j=1}^{Q^-} \lambda_j^- \sigma_{t-j}^{2-} \quad (14d)$$

$$\sigma_t^+ \sigma_t^- = \gamma^\mp + \sum_{i=1}^{K^\mp} \phi_j^\mp u_{t-i}^+ u_{t-i}^- + \sum_{j=1}^{Q^\mp} \lambda_j^\mp \sigma_{t-j}^+ \sigma_{t-j}^- \quad (14e)$$

Making use of a system of equations is required for efficiently estimating the parameters in the model since positive and negative components are not independent of each other. The variables $\sigma_t^{2+}$ and $\sigma_t^{2-}$ represent a measure of time-dependent conditional variance for $u_t^+$ and $u_t^-$, respectively, and $\sigma_t^+ \sigma_t^-$ is their conditional covariance. The symbols $\gamma$, $\Phi$ and $\lambda$ represent volatility parameters to be estimated for each case. The optimal lag orders $K^+$, $Q^+$, $K^-$, $Q^-$, $K^\mp$ and $Q^\mp$ can be determined by minimizing an information criterion. Model (14) can be estimated by using the multivariate GARCH method[4], which can capture the clustering property of volatility that usually characterizes financial assets.[5] The model can also be estimated by making use of the multivariate t-distributions in order to remedy the potential fat-tails effect to some extent. Since our proposed approach emphasizes two hedge ratio parameters, i.e., $h^+$ and $h^-$, it is crucial to assess not only if these coefficients are statistically significant individually, but also if their difference $(h^+ - h^-)$ is statistically significant. Thus, the following null hypothesis can be tested by the Wald (1949) test: $H_0: h^+ - h^- = 0$ against the alternative $H_1$:

---

[4] For a survey on multivariate GARCH models see Silvennoinen and Teräsvirta (2009).
[5] However, this issue can be also settled empirically by testing the null hypothesis of constant variance against the time-varying conditional variance by conducting Hacker and Hatemi-J (2005) multivariate ARCH test. If the null is not rejected, then equations (14a) and (14b) can be estimated via the seemingly unrelated regression equations (SURE) method. Other alternative methods for estimating the system of regression equations can also be considered such as the multivariate GMM, the three stage least squares, or the full information maximum likelihood method.

$h^+ - h^- \neq 0$. If the underlying null hypothesis is not supported empirically, it can be concluded that the optimal hedge ratios are asymmetric. In that case, the investor with a short position in the asset needs to make use of $h^+$ for hedging purposes and the investor that is long should rely on $h^-$.

## 4. An Application

The suggested asymmetric test is applied to estimating the optimal hedge ratio for Bitcoin. The sample period covers the last week of December in 2017 until the first week of March in 2024. The spot and futures exchange rates of Bitcoin against the US dollar are used. The choice of the start of the sample period is restricted by the data availability since Bitcoin futures were launched by CBOE (the Chicago Board Options Exchange) and CME (the Chicago Mercantile Exchange) in December 2017 as an instrument for hedging the price risk of Bitcoin.[6] The source of the data is Yahoo Finance. The estimation results based on the system of equations (14), which is estimated via a multivariate GARCH(1, 1) model assuming a multivariate t-distribution, are presented in Table 2.[7]

Table 2. Estimation Results.

| Parameters | Estimated Parameter Values |
|---|---|
| $h^+$ | 0.399432 |
| $h^-$ | 0.713761 |
| | |
| Null Hypothesis | P-value |
| $H_0: h^+ = 0$ | < 0.00001 |
| $H_0: h^- = 0$ | < 0.00001 |
| $H_0: h^+ - h^- = 0$ | < 0.00001 |

As it is evident from the estimation results, each individual optimal hedge ratio is statistically significant strongly. The null hypothesis that the two hedge ratios are equal to each other is also rejected strongly. Thus, the optimal hedge ratio for Bitcoin is indeed asymmetric. This in turn implies that the trader who is short in Bitcoin needs to have a long position of 0.399432 unit in the futures for each unit of the spot for optimal hedging. On the other hand, the trader that is long in Bitcoin should have a short position of 0.713761 unit in the futures for each unit of the spot for ideal hedging. Surprisingly, the optimal hedge ratio for a long position in Bitcoin is 78.69% higher than the optimal hedge ratio for a trader with a short position in the underlying asset.

---

[6] According to Köchling et. al., (2019) the information efficiency of Bitcoin has increased since the introduction of Bitcoin futures. Similar effects on other major cryptocurrencies were not found by the mentioned the authors, however.
[7] Here the focus is on the estimated values for the optimal hedge ratios only. Thus, the rest of the estimated parameters are not presented.

## 5. Conclusions

Hedging is a crucial strategy utilized by investors and enterprises for reducing risk. This paper proposes an efficient methodological framework for testing whether or not the optimal hedge ratio is asymmetric. In case the null hypothesis of symmetry is rejected, each position dependent hedge ratio can be obtained via the suggested model. An application is provided using the spot and futures prices of Bitcoin. The results show that the null hypothesis of symmetric optimal hedge ratio can be strongly rejected. The estimation results reveal also that the optimal hedge ratio for a long position in Bitcoin is almost 78% higher compared to the optimal hedge ratio for a short position in this very risky asset. Thus, using the same hedge ratio for both positions, as it is common practice in the existing literature (see Corbet et. al., 2018; Sebastião and Godinho 2020; and Matsui and Knottenbelt 2023, among others), can be an inefficient strategy for reducing risk in this case. However, in order to establish the robustness of the obtained empirical evidence further empirical investigations with regards to other assets and different time horizons are needed in the future using the suggested approach.

**Acknowledgements**

The current research is partly financed by the CBE Annual Research Program (CARP) 2024 funded by the United Arab Emirates University. The usual disclaimer applies, however.